\documentclass[%
 aip,
 amsmath,amssymb,
apl,
reprint,%
]{revtex4-1}
\usepackage{graphicx}
\usepackage[percent]{overpic} 
\usepackage{color}
\usepackage{dcolumn}
\usepackage{bm}
\usepackage{epstopdf}

\newcommand{\micron}{$\mu$m}
\newcommand{\Ohm}{$\Omega$}

\begin{document}

\preprint{AIP/123-QED}

\title[]{Photon-noise limited sensitivity in titanium nitride kinetic inductance detectors}

\author{J.Hubmayr}
\email{hubmayr@nist.gov}
\affiliation{National Institute of Standards and Technology, 325 Broadway, Boulder, CO 80305, USA}
\author{J.~Beall}
\affiliation{National Institute of Standards and Technology, 325 Broadway, Boulder, CO 80305, USA}
\author{D.~Becker}
\affiliation{National Institute of Standards and Technology, 325 Broadway, Boulder, CO 80305, USA}
\author{H.-M.~Cho}
\affiliation{National Institute of Standards and Technology, 325 Broadway, Boulder, CO 80305, USA}
\author{M.~Devlin} 
\affiliation{University of Pennsylvania, Department of Physics \& Astronomy, 209 South 33rd St, Philadelphia, PA 19104, USA}
\author{B.~Dober}
\affiliation{University of Pennsylvania, Department of Physics \& Astronomy, 209 South 33rd St, Philadelphia, PA 19104, USA}
\author{C.~Groppi}
\affiliation{Arizona State University, School of Earth \& Space Exploration, 781 S Terrace Rd, Tempe, AZ 85281, USA}
\author{G.C.~Hilton}
\affiliation{National Institute of Standards and Technology, 325 Broadway, Boulder, CO 80305, USA}
\author{K.D.~Irwin}
\affiliation{Stanford University, Department of Physics, Stanford, CA 94305, USA}
\author{D.~Li}
\affiliation{National Institute of Standards and Technology, 325 Broadway, Boulder, CO 80305, USA}
\author{P.~Mauskopf}
\affiliation{Arizona State University, School of Earth \& Space Exploration, 781 S Terrace Rd, Tempe, AZ 85281, USA}
\author{D.P.~Pappas}
\affiliation{National Institute of Standards and Technology, 325 Broadway, Boulder, CO 80305, USA}
\author{J.~Van Lanen}
\affiliation{National Institute of Standards and Technology, 325 Broadway, Boulder, CO 80305, USA}
\author{M.R.~Vissers}
\affiliation{National Institute of Standards and Technology, 325 Broadway, Boulder, CO 80305, USA}
\author{Y.~Wang}
\affiliation{National Institute of Standards and Technology, 325 Broadway, Boulder, CO 80305, USA}
\affiliation{Southwest Jiaotong University, Quantum Optoelectronics Laboratory, Chengdu, China}
\author{L.F.~Wei}
\affiliation{Southwest Jiaotong University, Quantum Optoelectronics Laboratory, Chengdu, China}
\author{J.~Gao}
\affiliation{National Institute of Standards and Technology, 325 Broadway, Boulder, CO 80305, USA}

\date{\today}


\begin{abstract}
We demonstrate photon-noise limited performance at sub-millimeter wavelengths in feedhorn-coupled, microwave kinetic 
inductance detectors (MKIDs) made of a TiN/Ti/TiN trilayer superconducting film, tuned to have a transition 
temperature of 1.4~K.  Micro-machining of the silicon-on-insulator wafer backside creates a quarter-wavelength backshort optimized for efficient coupling 
at 250~\micron.  Using frequency read out	 and when viewing a variable temperature blackbody source, we measure device 
noise consistent with photon noise when the incident optical power is $>$~0.5~pW, corresponding to noise equivalent powers 
$>$~3$\times 10^{-17}$ W/$\sqrt{\mathrm{Hz}}$.  This sensitivity makes these devices suitable for broadband photometric applications 
at these wavelengths.    
\end{abstract}

\maketitle

Microwave kinetic inductance detectors (MKIDs) are superconducting pair breaking devices \cite{day2003} currently 
in development for a broad range of applications \cite{baselmans2012a}.  Making use of high quality factor resonators, MKIDs multiplex in the frequency domain and thus scale 
to large-format arrays.  MKIDs are the implemented or planned detector technology 
for several sub-millimeter instruments \cite{golwala2012a,swenson2012a,shirokoff2012a,monfardini2013a,hubmayr2014dual}.
  
The fundamental limit of the sensitivity of a sub-millimeter photon integrating detector is set by the photon fluctuations from the source under
observation, which is referred to as the photon-noise limit or background limit.  The noise equivalent power (NEP) of
photon fluctuations from a narrow bandwidth source can be expressed as \cite{zmuidzinas2003a} 
\begin{eqnarray}
\label{eqn:photon_nep}
NEP_{photon}~=~\sqrt{\frac{2Ph\nu(1+m\eta)}{\eta}}.
\end{eqnarray}
Here and throughout this letter, we refer NEP terms to the input of an optical system with optical efficiency $\eta$. The first term describes Poisson shot noise.  $P$ is the input optical power, $h$ is Planck's constant, and $\nu$ is the 
observation center frequency.  The second term describes photon bunching, where $m$ is the occupation number per mode.  MKID detectors are also limited by generation-recombination noise \cite{deVisser2011a,wilson2001a}.  
In the limit of photon dominated quasiparticle production, only the recombination noise need be considered, and the NEP can be expressed as \cite{yates2011a}
\begin{eqnarray}
\label{eqn:gr_nep}
NEP_R~=~\sqrt{\frac{2P\Delta/\eta_{pb}}{\eta}}.
\end{eqnarray}
Here $\Delta$ is the superconducting energy gap and $\eta_{pb} \lesssim 0.57$ is the efficiency of converting photons to quasiparticles \cite{guruswamy2014}.
In this work since $h\nu \sim 22 \Delta$, the recombination noise is sub-dominant to the photon-noise.   

Recently, the sensitivity of MKIDs has improved.  Laboratory studies have shown photon-noise limited performance in aluminum MKIDs \cite{devisser2014a, janssen2013a} as well 
as TiN-based MKIDs\cite{mckenney2012a}.  As the superconducting material in 
kinetic inductance detectors, TiN has a number of potential 
advantages including low loss \cite{leDuc2010a,vissers2010a}, high resistivity, large kinetic inductance fraction and a tunable transition temperature.  
The spatial non-uniformity of TiN films with $T_c~<~4~K$ has been improved by use of a new superconducting film, a TiN/Ti/TiN trilayer \cite{vissers2013a}.  
These films show $<$~1\% $T_c$ variation across a 75~mm diameter wafer.

\begin{figure*}[t]
\begin{center}
\includegraphics[width=\textwidth]{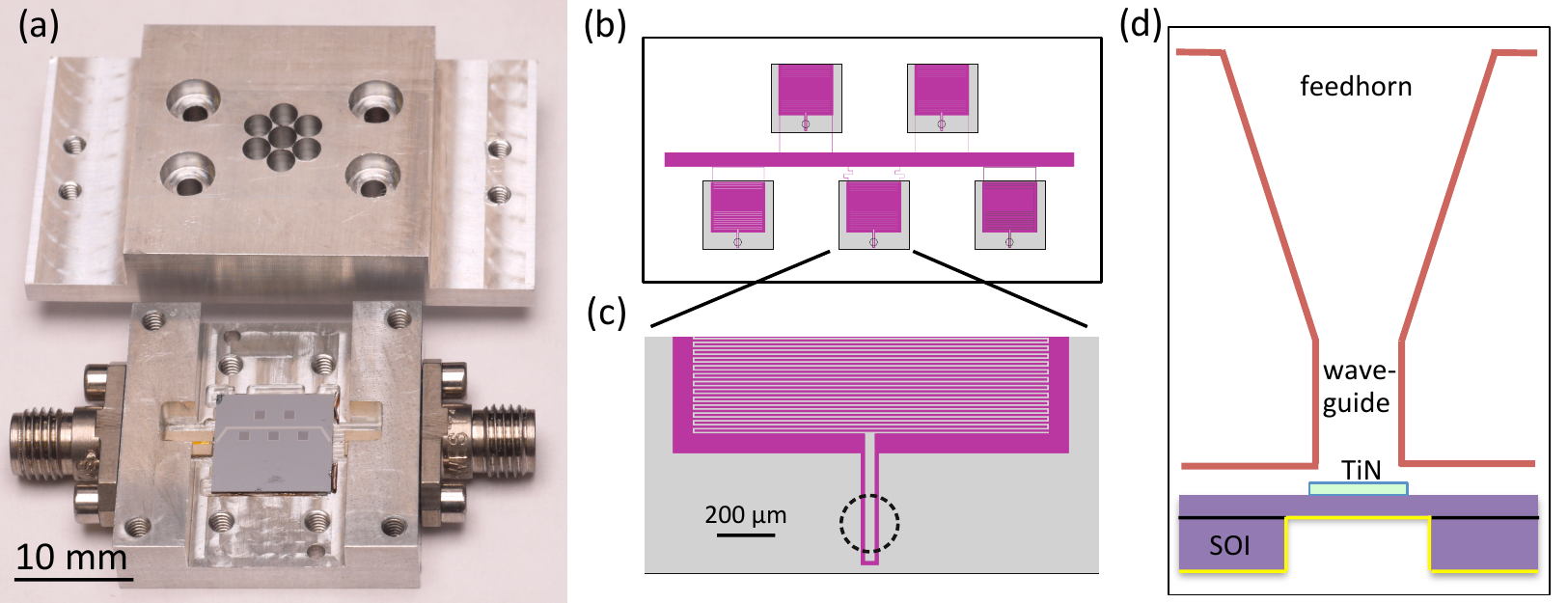}
\caption{Detector package and optical coupling. {\bf (a)} Photograph of 5-pixel die mounted in a sample holder together with the feedhorn array.  {\bf (b)} Detector chip design showing that the cross-wafer microstrip feedline couples to five lumped-element resonators. The grey squares 
indicate where backside silicon is removed from the SOI wafer to create the quarter-wavelength backshort.  {\bf (c)} Zoom-in view of 
a single pixel comprised of a 5~\micron\ IDC and single turn inductor.  The dashed circle locates the position of the feedhorn's exit waveguide, the optically 
active region. {\bf (d)} Cross-section schematic (not to scale) of the feedhorn/waveguide fed coupling scheme, which illustrates the silicon membrane backshort.}
\label{fig:sample}
\end{center}
\end{figure*}

In this work, we report a measurement of photon-noise limited sensitivity at 250~\micron\ in kinetic inductance detectors fabricated from TiN/Ti/TiN films at 
power loads relevant for balloon-borne or satellite-based photometry ($1<P<20$~pW).  These polarization-sensitive detectors are under development for 
the next-generation BLAST experiment \cite{galitzki2014next}.  The 1.4~K transition temperature of the film is chosen to accommodate 
operation from a 300~mK bath temperature, as is planned for BLAST.  The devices are feedhorn-coupled to a variable temperature 
blackbody source.  Feedhorn-coupling is a standard approach at sub-millimeter wavelengths that has recently been demonstrated with MKIDs \cite{mccarrick2014horn}.  Our measurements show traits associated with photon-noise limited detection in MKIDs.  The NEP scales as $\sqrt{P}$, and the noise spectra are flat with a roll-off set by quasiparticle recombination and 
resonator ring-down time.

Figure~\ref{fig:sample} shows the detector array package, detector design and feedhorn-coupled mounting scheme.  
We grow a 4/10/4~nm thick trilayer film of TiN/Ti/TiN as described in \cite{vissers2013b} on a 
silicon-on-insulator (SOI) wafer.  The film is patterned into five $f\sim1$~GHz lumped-element resonators 
each comprised of a 5~\micron\ width and spacing interdigitated capacitor (IDC) of total area 0.9~mm$^2$ in parallel with a 
one-turn inductor.  The width of the inductor strip is 8~\micron, and has a total volume $V$~=~86~\micron$^3$.  
The resonators on the chip couple to a 50~\Ohm\ cross-wafer microstrip feedline, also made of the 
trilayer.  We measure $T_c$~=~1.4~K, internal quality factors of 200,000 to 400,000 at bath temperature $T_{bath}$~=~75~mK, and coupling quality factors $\sim$~30,000.

The inductor element also acts as the absorber of incident sub-millimeter radiation. It is located $\sim$~50~\micron\ below the 173.7~$\pm$~1~\micron\ diameter waveguide 
output of the feedhorn.  The sheet impedance 
(90~$\Omega/\Box$) of the inductor is matched to the waveguide impedance and absorbs radiation polarized along the long axis of the inductor.   
The silicon behind the inductor and capacitor is removed with a deep reactive ion etch up to the buried 
insulator layer of the SOI wafer.  This produces a 19~\micron\ thick silicon membrane.  The oxide layer is removed with 
a CHF$_3$/O$_2$ plasma etch.  RF sputter deposition of a 500~nm thick layer of Nb on the 
backside of the wafer creates a quarter-wavelength reflective backshort as well as a continuous ground plane.

\begin{figure*}[t!]
\begin{overpic}[width=\textwidth]{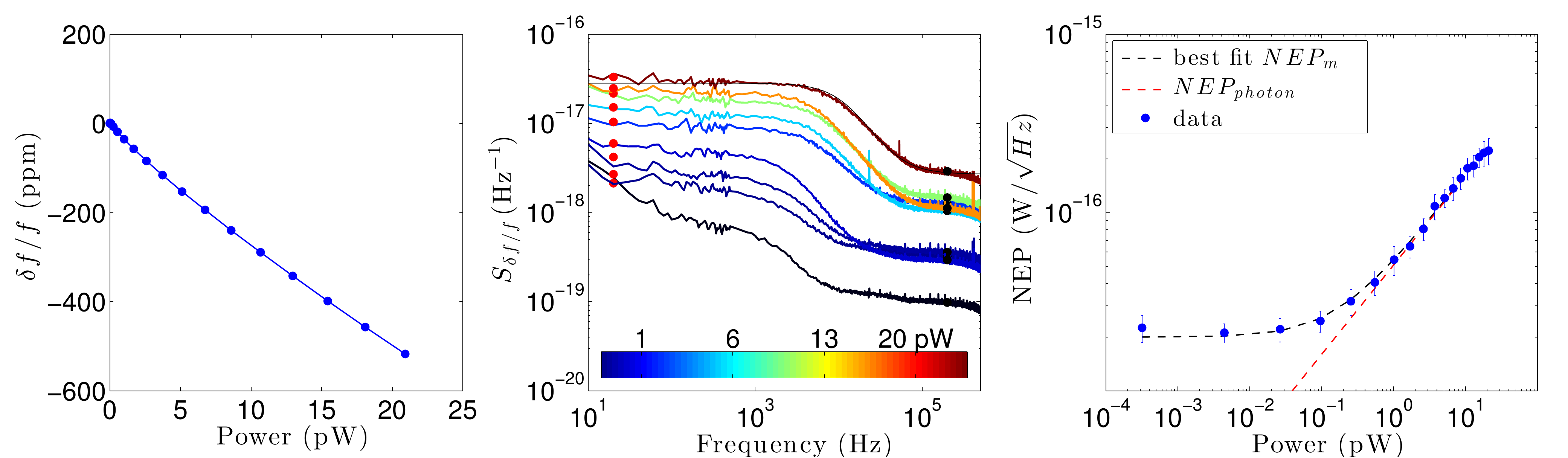}
\put(16,28){\normalsize \textcolor{black}{\bf (a)}}
\put(48,28){\normalsize \textcolor{black}{\bf (b)}}
\put(82,28){\normalsize \textcolor{black}{\bf (c)}}
\end{overpic}
\caption{{\bf (a)} Fractional frequency response of detector to thermal power emitted by the blackbody load. {\bf (b)} Noise spectra, in fractional frequency units, 
for blackbody loads between 4~K and 21~K taken at a bath temperature of 75~mK.  
The solid line is an example Lorentzian model fit to the $P~=~20$~pW data.  {\bf (c)} NEP versus thermal load.  The low frequency noise (20 Hz, red points) from the 
spectra in panel (b), corrected for system noise using the 200 kHz black points, is 
converted to NEP by use of the local slope in (a).  
These data 
are in good agreement with the photon-noise prediction (the red, dashed line) at power levels $>$0.5~pW.  The black, dashed line is the best fit NEP model, which includes contributions due to photon noise, recombination noise, and a noise term independent of optical power.}
\label{fig:NEPsummary}
\end{figure*}

We mount this package to the 50~mK stage of an ADR cryostat.  The feedhorns view a temperature 
controlled THz tessellating tile that has $<-30$~dB reflection at 600~GHz \cite{thz_tile_performance}.  Using the simulated beam profile of the feedhorns, we calculate $>$~99\% of the optical 
throughput goes to the blackbody.  The tile is glued into a copper block that is weakly linked to the 3~K stage of the cryostat.  We control the load 
temperature between 4~K and 21~K to $\sim$~1~mK stability by use of a heater and calibrated thermometer.  

The optical passband is defined by the $\nu_1$~=~1~THz waveguide cut-off in the feedhorn and a well-characterized 1.4~THz low-pass filter \cite{pascale2008a} mounted to the 
feedhorn array.  The in-band power emitted from the load is,
\begin{eqnarray}
\label{eqn:TtoP}
P = \int_{\nu}d\nu \left (\frac{c}{\nu} \right ) ^2 B(\nu,T) F(\nu),
\end{eqnarray}
where we have assumed single-mode optical throughput.  $B(\nu,T)$ is the Planck function.  The passband $F(\nu)$ includes the measured filter transmission and a 
calculation of the waveguide cut-off.  
Metal-mesh filters are known to have harmonic leaks at frequencies above cut-off \cite{ade2006review}.  However, the integrated power above the passband is $<$~2\% of the total in-band power, even at the highest blackbody temperatures.  

We perform a frequency sweep and characterize noise with a homodyne measurement and a 
SiGe amplifier at thermal loads ranging from $P$~=~0.3~fW to 21~pW and at $T_{bath}$~=~75~mK.  A fit to the complex transmission S$_{21}(f)$ yields the resonant frequency as a function of thermal power.    
The fractional frequency shift $\delta{f}/f$ response of a detector to the changing thermal load is shown in Fig.~\ref{fig:NEPsummary}a.  In a separate experiment, we confirm there is no stray coupling to the thermal 
source by blocking all 
horn apertures with reflective tape and observing no frequency shift to applied thermal load.

For each thermal load, we measure noise at the microwave frequency that maximizes $\delta{S_{21}}/\delta{f}$.  
The power on the feedline is in a range -97 dBm $\lesssim P_g \lesssim$ -87 dBm, chosen to be at least 3~dB below bifurcation to the strong 
non-linear effects in the resonator \cite{zmuidzinas2012a,swenson2013a}.  We project the raw in-phase and quadrature components of the data into the frequency and dissipation quadratures \cite{zmuidzinas2012a} 
and examine the noise in the frequency quadrature.  

The spectrum of a photon-noise limited MKID is expected to be of Lorentzian form.  
Example spectra at various thermal loads are shown in 
Fig.~\ref{fig:NEPsummary}b.  
When $P~>$~0.5~pW, each spectrum is well described by a Lorentzian function of white noise level $A$ and time constant $\tau$ 
that is summed with a background noise floor $B$,   
\begin{eqnarray}
\label{eqn:noisefit}
S_{\delta{f}/f}(\omega)= \frac{A}{1+\omega^2 \tau^2}+B.
\end{eqnarray}
An example fit to the $P~=~20~$~pW spectrum is shown as the solid, black line in Fig.~\ref{fig:NEPsummary}b.  We observe 
that the white noise level increases with applied thermal power. 

The detector NEP may be determined by dividing the square root of the detector noise in Fig.~\ref{fig:NEPsummary}b by the local 
responsivity $\frac{\delta{f}/f}{\delta{P}}(P)$ in Fig.~\ref{fig:NEPsummary}a.  
We estimate the detector noise by subtracting the amplifier noise contribution (black dots at 200~kHz in Fig.~\ref{fig:NEPsummary}b) 
from the average noise at 20~Hz(red dots in Fig.~\ref{fig:NEPsummary}b), which is determined by a power law fit to the data around 20~Hz.   
These values, together with one standard deviation fit uncertainty, are plotted as a function of thermal load power $P$ in Fig.~\ref{fig:NEPsummary}c, which is the main result of this letter.  
Above 0.5~pW the measured 
NEP scales as $\sqrt{P}$, as expected for a photon-noise limited detector.  Furthermore, the red, dashed line of Fig.~\ref{fig:NEPsummary}c 
is the photon noise prediction, 
calculated using Eq.~\ref{eqn:photon_nep} ($\nu$~=~1.25~THz, $\eta$~=~0.69 and $m~<~0.1$ for all blackbody temperatures).  For $P~\geq$~0.5~pW, the data are in good agreement with the photon-noise prediction. 

The full data set fits the model (dashed-black line of Fig.~\ref{fig:NEPsummary}c) 
\begin{eqnarray}
NEP_m^2 = NEP_{\alpha}^2 + \frac{NEP_{photon}^2+NEP_R^2}{\eta},
\end{eqnarray}
which has been used to determine the optical efficiency $\eta$ \cite{yates2011a} .  ($NEP_{\alpha}$ is a constant term, independent of optical power and discussed below).
The detector responds to both co- and cross-polar radiation as $\delta{f} \propto \eta_c P_c + \eta_x P_x$, where $\eta_c (\eta_x)$ and $P_c (P_x)$ are the 
efficiency and power emitted in the co- (cross-) polar component respectively.  The frequency responsivity and NEP we report are referred to the power 
in a single polarization from a source with equal power in both polarizations, as is the case for the thermal source used in our experiment.  As a result of this description, $\eta = \eta_c + \eta_x$.  
The fit yields $\eta=0.69 \pm 0.01$, which is lower than the sum $\eta_c + \eta_x$=0.64+0.16=0.80 determined from electromagnetic coupling simulations, which assume a $\Delta{Z}$~=~50~\micron\ air gap between the absorber and 
waveguide.  The discrepancy may be due to uncertainty in the positioning of the absorber with respect to the waveguide output 
or absorption in the aluminum feedhorn, which has not been accounted for.  
To pin down $\eta_c$ and $\eta_x$ separately requires measurements with a polarizer.  The design and measured polarization performance of this detector type 
is the subject of a future publication.  For polarimetry applications such as BLAST \cite{galitzki2014next}, $\eta_c \gg \eta_x$ is required.  By splitting the absorber width into 2~\micron\ strips, simulations suggest that $\eta_c \sim 0.8$ and $\eta_x \lesssim 0.02$ are achievable.   

At $P<0.1$~pW, we find that the basic shape of the spectra are unchanged and that the NEP saturates to $NEP_{\alpha}$~=~2~$\times$~10$^{-17}$~W/$\sqrt{\mathrm{Hz}}$.  
Initial measurements suggest that the source of this noise is from a background of excess quasiparticles due to stray light absorption.  
We will conduct future tests in a carefully designed light-tight box to confirm that $NEP_{\alpha}$ decreases when stray light is 
reduced.   However, 
the current value of $NEP_{\alpha}$ has a small effect on the sensitivity for many potential applications.

Lastly, we discuss the implications of the near constant responsivity to increased thermal loading (Fig.~\ref{fig:NEPsummary}a).  This phenomenon has previously been 
observed in TiN films \cite{mckenney2012a,noroozian_thesis} and departs from 
the behavior in conventional superconducting materials, which show $\delta{f}/\delta{P} \sim P^{-0.5}$ as a consequence of 1) $\delta{f} \propto \delta{n_{qp}}$ and 
2) $P \propto n_{qp}^2$.  Here $n_{qp}$ is the quasiparticle number density.   
Our linear responsivity measurement suggests that 1), 2) or both relationships are not valid for our TiN trilayer films.  If $\delta{f} \propto \delta{n_{qp}}$ holds, the 
observed constant responsivitiy 
suggests $P~\propto~n_{qp}$, which further implies that the quasiparticle lifetime time $\tau_{qp}$ is independent of $P$.  
However, the lifetimes inferred from the increasing roll-off in the noise is inconsistent with this picture, which may indicate that relationship 1) does not hold 
for this material.  This apparent discrepancy remains outstanding and 
requires future investigation.  Understanding the recombination physics of TiN films is key to explaining the observed 
linear responsivity.  We plan to investigate this device physics by directly measuring the detector response to optical pulses in future experiments.  

In conclusion, we have demonstrated photon-noise limited sensitivity in feedhorn-coupled 
microwave kinetic inductance detectors fabricated from a superconducting TiN/Ti/TiN trilayer film.  This work represents a significant step 
towards realizing high sensitivity, large-format kinetic inductance detector arrays suitable for broadband photometry 
and with polarimetric sensing capability.  Further work is planned to explore the origin of $NEP_{\alpha}$ and to better understand 
the anomalous quasiparticle physics in our TiN/Ti/TiN films.

This work was supported in part by NASA through grant number NNX13AE50G S04. 
TiN materials research is supported in part by DARPA.  Brad Dober is supported by the NASA Earth and Space Science Fellowship.  
The authors would like to thank Edward Wollack and Lev Ioffe for useful discussions.



\end{document}